# Model-based Cybersecurity Analysis: Past Work and Future Directions

Simon Yusuf Enoch, The University of Queensland
Mengmeng Ge, Deakin University
Jin B. Hong, The University of Western Australia
Dong Seong Kim, The University of Queensland

*SUMMARY & CONCLUSIONS*

Model-based evaluation in cybersecurity has a long history. Attack Graphs (AGs) and Attack Trees (ATs) were the earlier developed graphical security models for cybersecurity analysis. However, they have limitations (e.g., scalability problem, state-space explosion problem, *etc.*) and lack the ability to capture other security features (e.g., countermeasures). To address the limitations and to cope with various security features, a graphical security model named attack countermeasure tree (ACT) was developed to perform security analysis by taking into account both attacks and countermeasures. In our research, we have developed different variants of a hierarchical graphical security model to solve the complexity, dynamicity, and scalability issues involved with security models in the security analysis of systems. In this paper, we summarize and classify security models into the following; graph-based, tree-based, and hybrid security models. We discuss the development of a hierarchical attack representation model (HARM) and different variants of the HARM, its applications, and usability in a variety of domains including the Internet of Things (IoT), Cloud, Software-Defined Networking, and Moving Target Defenses. Moreover, we discuss the pros and cons of each variant of HARM based on its applications and usage. Furthermore, several security metrics have been developed to be used with the graphical security model (including HARMs) to analyze the security posture of the systems and evaluate the effectiveness of defense mechanisms which is also being taken as input into optimization algorithms to compute optimal defense deployment. Thus, we provide the classification of the security metrics, including their discussions. Finally, we highlight existing problems and suggest future research directions in the area of graphical security models and applications. As a result of this work, a decision-maker can understand which type of HARM will suit their network or security analysis requirements.

## 1 INTRODUCTION

Over the past years, networks have become complex and dynamic, connecting different components and applications. This has introduced numerous relationships between the interconnected systems and applications. While these advancements have brought a lot of benefits to our daily lives in terms of file storage, improved communications, networking, *etc.,* cyber-attackers can find exploitable vulnerabilities on critical systems and compromise them to take full control or cause further damage. Hence, it is important to identify these vulnerabilities and mitigate them. In-depth security modeling and analysis can assess security vulnerabilities and identify the relationship between the vulnerabilities, which can be effectively protected with appropriate defense strategies.

Graphical Security Modeling (GSM) has been the widely adopted method to model and analyze vulnerabilities, cybersecurity events, and to quantify security based on the structure of the models. Moreover, possible defense strategies can be evaluated and analyzed with GSM along with security metrics. Attack Graph (AG) [1, 2] and Attack Trees (ATs) [3] are the most common type of GSMs. The AG shows potential sequences of attacker's steps by enumerating all potential attack paths that an adversary can use to penetrate a networked system, however, with the increasing complexity and dynamicity of modern networks, the AG has exponential complexity which causes scalability problem. On the other hand, the ATs represent attacks as a tree with leaf nodes and child nodes, where leaf nodes show different ways of achieving the goal, and child nodes represent specific attack actions. However, the AT does not explicitly reflect the sequences of attackers' actions, and its formalism does not capture countermeasures. Defense trees [4] are ATs with countermeasures but they place countermeasures only at the leaf nodes.

To incorporate countermeasures at both the leaf nodes and intermediates nodes of ATs, and to also avoid the state-space explosion problem, Roy *et al*. [5] proposed the Attack Countermeasures Tree (ACT) for security analysis. While Hong and Kim addressed the scalability problem associated with GSMs [6, 7] by developing hierarchical graphical security models that combine the AGs and ATs into two or more layers. This model is named Hierarchical Attack Representation Models (HARM). A two-layer HARM compromises of two layers: the upper layer which captures the network reachability information and the lower layer that captures the vulnerability information of each node in the network.

Furthermore, due to the dynamicity of the cloud, it is essential to extend the capabilities of the HARM. As a result, T-HARM [8] has been proposed to capture temporal states of the cloud configurations at different times to evaluate dynamically changing security posture. Further, a tool named *CloudSafe* [9] has been developed, which can be adopted by cloud service providers to evaluate security. Moreover, dynamic security metrics have been developed to evaluate the changing security postures. Existing security metrics, such as the probability of attack success, risk, and Return on Investment (RoI), are not suitable to reflect the system changes, which may lead to insecure network states that can be exploited by the attackers. Therefore, dynamic security metrics capture the security postures of all network states over the time period, providing a global view of the security posture changes. Hence, these metrics also allow global security optimizations in the cloud using the T-HARM.

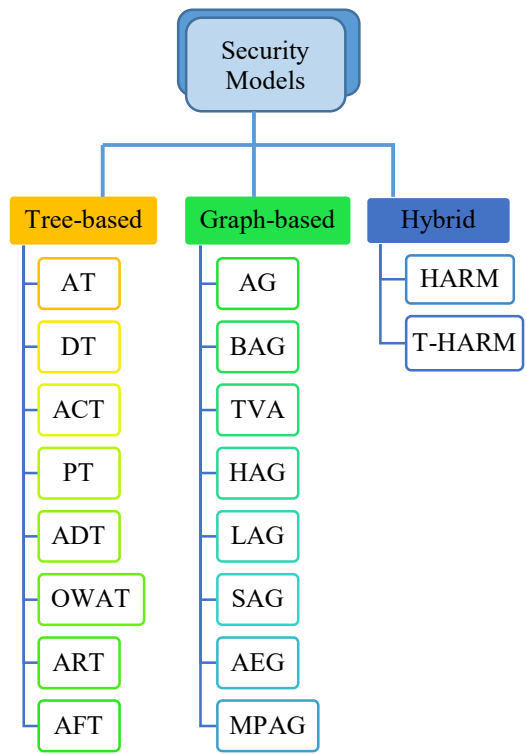

Figure 1: The variants of GSMs

The hierarchical graphical security model based on HARM has been proposed to automate security assessment for the IoT [10]. Potential attack paths are captured to depict sequences of attack actions and security metrics are developed to evaluate the security level of the IoT. Due to constrained resources and limited computation capabilities, several proactive defense mechanisms have been proposed, including MTD and cyber-deception. The hierarchical graphical security model has also been applied in evaluating the effectiveness of these defense mechanisms thus being used as input of optimization algorithms to compute optimal defense deployment. The contribution of this paper is summarized as follows.

- Survey the development of graphical security models (e.g., AGs, ATs, ACTs, and HARMs).
- Discuss the usability and applications of the variants of Hierarchical Attack Representation Models (HARMs), including their pros and cons.
- Classify security metrics based on usability and applications.
- Discuss future research directions in terms of security models, evaluations, and applications.

The rest of this paper is organized as follows. Section 2 provides the background of model-based security evaluation and analysis. Section 3 presents the ACTs and its evaluation method. In addition, the HARMs and their development over time are also presented, including the development of security metrics for the HARMs analysis. In Section 4, we present the applications of the different variants of HARMs in terms of network types, defense mechanisms, measurements, simulations, and experiments. Finally, Section 5 summarizes our paper and discusses future research directions.

## 2 BACKGROUND

Model-based security evaluation provides a systematic way to capture possible attack scenarios and analyze security based on system vulnerabilities. The GSMs have gained a lot of attention from security researchers [11] and industries [12]. In this section, we provide a brief background to the GSM-based security evaluation.

We group the GSMs into three categories; Graph-based, Tree-based, and Hybrid models. The survey in [13, 14] describes the whole family of security models with their various capabilities. The graph-based models found their origins from the concept of a privilege graph introduced by Dacier and Deswarte [15]. The privilege graph is a directed graph with nodes that represent privileges, and edges that represent vulnerabilities. The tree-based model is attributed to Weiss [16] who proposed the threat logic tree and later it was extended by Salter *et al*. [17] with threats countermeasures in the form of ATs. Hong and Kim [6] developed the hybrid model to improve the scalability of security models, where both the graph-based and tree-based models are used in different layers.

Over the years, different variants of the graph-based, tree-based, and hybrid models have been developed for various applications. In Figure 1, we show the different types of the graph-based, tree-based, and hybrid models with Protection Tree (PT) [18], Attack DT (ADT) [19], OWA tree (OWAT) [20], Attack Fault Tree (AFT) [21], Bayesian AG (BAG) [22], Topological Vulnerability Analysis (TVA) [23], Hierarchical AG (HAG) [24], Logical AG (LAG) [25], Security Augment Graph (SAG) [26], Attack Execution Graph (AEG) [27], Multiple Prerequisite AG (MPAG) [28], *etc*.

## 3 MODEL-BASED EVALUATION AND ANALYSIS

The AG and ATs provide the main platform for many security models, including the ACT and the HARM. In this section, we discuss the ACT and the HARMs, including its analysis over the years.

evaluate emerging networking technologies, such as cloud computing [31], Software-Defined Networking (SDN) [32], IoT [33], various variants of the HARM have been developed to adapt to the unique features of those networks and to assess their security. In Figure 2, we show the evolutions of the

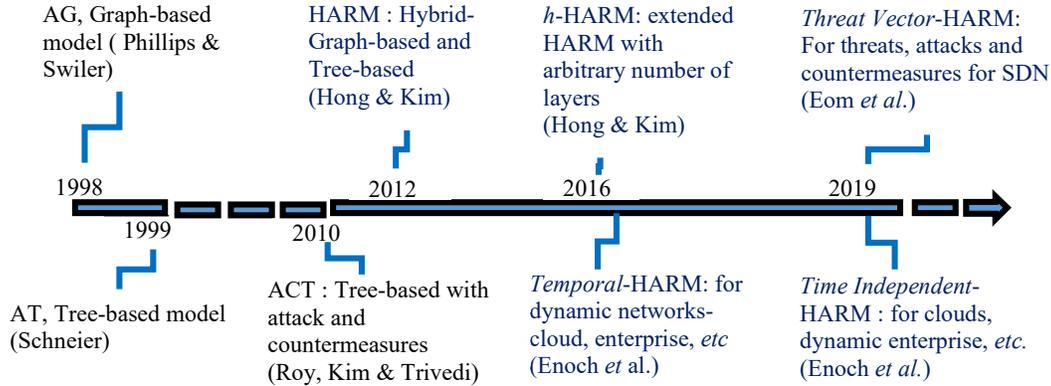

Figure 2: The HARMs' development over time

### 3.1 Attack Countermeasure Tree (Kishor's contributions)

Both AT and AG capture attack actions but do not include defense mechanisms. On the other hand, Defense Trees [4] capture the defense mechanisms in ATs but only at the leaf nodes. In Attack Response Trees (ART) [29], attacks and responses are included in any node based on a partially observable stochastic game model; however, the model suffered from a state-space explosion. To address these challenges, Roy, Kim & Trivedi developed the Attack Countermeasure Trees for modeling of attacks, defenses, and cybersecurity analysis [5, 30]. The ACT was first introduced in [5] and then further extended in [30] with quantitative analysis and optimization. The ACT is an extended defense tree that places detections and mitigations at the leaf node and the intermediate nodes of the tree. It contains three main events: attack events, detection events, and mitigation events. Using the ACT, security of a network can be analyzed and evaluated in terms of minimal cut sets, attack and security investment cost, Birnbaum importance measure, system risk, the impact of an attack, RoI, and Return on Attack (RoA). Moreover, both attacks and countermeasures can be prioritized based on the structural and Birnbaum importance measures in the ACT.

Furthermore, due to constraints and challenges in finding optimal countermeasures from a pool of countermeasures, the ACT has been used with single or multi-objective optimization techniques to compute and evaluate suitable countermeasures for optimal security. Besides, this addresses the problem of the state-space explosion in ART.

### 3.2 The HARM and its developments

In this section, we introduce the HARM and its developments over the years.

Due to the lack of effective techniques to assess and development of HARM to deal with concerns of various network technologies.

In 2012, Hong and Kim [6] introduced a two-layered hierarchical model ( HARM) to address the scalability problem of a single-layered security model (e.g., the AG, AT, etc.). In particular, Hong and Kim combined the AG and the AT in the same model but on a different layer to improve the scalability and reduce the computational complexity of GSMs, where the model constructions in the layers are independent of each other. Hence, this improves the HARM computational complexity, and evaluation compared to a single-layer AG. Besides, potential attack paths are explicitly captured in the upper layer of the HARM which cannot be captured in the ATs.

*h*-HARM: To further improve the scalability of HARM, Hong and Kim [7] developed h-layered HARM, where h represents an arbitrary number of layers with each layer having its security model that performs a separate functionality. For example, in [7] a three-layer HARM for a network is shown, where the upper layer captures the relationship between the network subnets using AG that models the reachability, the middle layer captures the reachability between the network hosts based on AG, and the lower layer captures the relationship between vulnerabilities of the hosts using ATs. This approach decreases the complexity of the security evaluation compared to the 2-layered HARM because the computations are performed in the different layers. So, using more layers with the h-HARM will further decrease the computational complexity.

**Pros and Cons:** The *h*-HARM has shown to be more scalable than the single-layered security model. Using the *h*-HARM potential attack paths can be captured and analyze in the model within a short time. In addition, the *h*-HARM has been adopted to analyze the security of different types of networks such as enterprise, IoT, Cloud, SDN, *etc*. However, the *h*-HARM does not take into account security changes over time. Further, the scalability and complexity problem of security models still exist when the network grows larger.

**T-HARM**: Changes in networks lead to changes of the attack surfaces thus affecting security analysis. However, the AG, AT, and HARM does not take into account the various changes that happen in the network. For instance, a network attack surface changes when a new host is connected to the network (e.g., bring your own device), an update of software vulnerabilities, the discovery of new vulnerabilities, firewall configuration and settings changed, etc. As a result, it is essential to extend the capabilities of the HARM to model and analyze the security of dynamic networks.

Temporal graphs [34, 35] were mainly used to model changes in the social network but have not been used in the context of graphical security models. In our work [8, 36], we extended the capabilities of the HARM to model dynamic networks based on the temporal graphs. Specifically, we developed a temporal graphical security model to capture and analyze the security of dynamic networks at every time $t$. The temporal HARM captures the security changes onto two layers at various times; the temporal network topology is captured at the upper layer using AGs and the vulnerability information for each node at the lower layer using a set of ATs. By doing so, the possible security of the network states is captured and analyzed at various times, thus showing the changes in the network states at every time t.

**Pros and Cons:** The T-HARM is used to model security changes in dynamic networks (e.g., Cloud, SDN) over a period of time $t$, where the potential attack paths for a period of time are captured. Based on the T-HARM the security can be observed for a period of time $t$. Moreover, the effectiveness of dynamic defense techniques such as MTD is also evaluated with the T-HARM. However, network changes can occur even more frequently and as a result, important network states can be skipped in a security analysis, which the T-HARM cannot take into account. Furthermore, the T-HARM is not capable of aggregating security information to generate the overall security posture of dynamic networks.

**Time Independent HARM** (TI-HARM): We developed a time-independent graphical security model [37, 38] that captures all potential attack scenarios of dynamic networks regardless of network states and time. The main idea of the time-independent security model is to model the security of dynamic networks by aggregating the security components of multiple states to form a single GSM taking into account multiple states, duration of each state, and the visibility of components (e.g., hosts and connections) in the states. By doing so, all the possible network components observed in various network states are captured and modeled. TI-HARM allows us to model all possible attack scenarios including ones carried out in multiple network states on a single GSM without having to look at multiple GSMs. Hence, the overall overview of the network security (using metrics) can be calculated without looking at the multiple metrics for every time $t$.

**Pros and Cons:** The TI-HARM provides a more comprehensive security analysis since all-important network components are taken into account compared to a single network state model, that ignores other changes afterward. Moreover, all potential attack scenarios over time are modeled on one security model. The con of this approach is that the model gets larger in size when a high threshold value is used to construct the model, which will require further analysis to remove less important nodes.

**Threat-Vector HARM**: Software-Defined Networking (SDN) is one of the emerging networking technologies that extend the capabilities of existing networks by providing various functionalities such as modification of network configurations in real-time. However, unlike the traditional network, applications and communication protocols used in the SDN controllers may expose vulnerabilities. As a result, a two-layered Threat Vector HARM (TV-HARM) was developed [39], which is also an extension of the HARM to capture threat-based attacks rather than individual attacks taking into account the threat vectors in SDN. The TV-HARM analyzes existing and emerging threat vectors in the SDN, which includes capturing dynamic changes of SDN, measuring and evaluating attack and defense scenarios, and showing the security posture through various security metrics based on the SDN.

**Pros and Cons:** The TV-HARM is useful to model security threats and their effects on the SDN network. However, its applicability remains unknown for other network types.

**HARM Visualization**: To visualize the HARM and the security analysis, a web-based application named Safeview [40] was developed. The Safeview provides a graphical user interface consisting of an upper layer and a lower layer. The upper layer is visualized using a force layout and the lower layer is visualized using a tree layout [40]. The Safeview uses Data-Driven Documents library for the visualization implementation and interacts with the HARM engine based on Apaches2, HTML, PHP, and JSON. Furthermore, to visualize topology changes in IoT networks and to highlight attack paths in HARM, including the attackers' interaction with decoy systems, another web-based visualization interface was implemented based on the new technologies [41]. Besides, the visualization for the upper layer of HARM was also implemented for the cloud-band model with MTD techniques [42].

**Pros and Cons:** The Safeview is useful to visualize the HARM with small to medium-size networks. However, as the network size grows larger, the size of the nodes in the visualization becomes smaller in the view interface thus becoming harder to read.

**Summary**: To address the scalability problem of single-layered GSMs, *h*-HARM was developed. To address the problem of static security analysis, the temporal graphs are incorporated into the HARM to create T-HARM and TI-HARM to analyze the security of dynamic n*etworks*. The most obvious advantage of the proposed approaches is that it will capture the network temporal change into two different layers at a different time, thereby, improving the adaptability of the current approaches. Further, it provides a way to analyze the security weaknesses of a dynamic network. Moreover, the TV-HARM model's various threats and the security posture of SDN with respect to identified threat vectors. As a result, the impact of different threats in SDN is measured. Moreover, the proposed approaches provided a way to evaluate the effectiveness of countermeasures, including proactive cyber-defenses.

The development of the HARM over the years has provided an approach for the high-level decision-makers to get an overview of the security of dynamic networks without the technical details and to understand appropriate defenses to deploy.

## 4 THE USABILITY AND APPLICATIONS OF THE HARMS

In this section, we discuss the security metrics that have been used with the HARMs to evaluate the security of enterprise networks, Cloud, SDN, IoT, and the effectiveness of defenses (preventive and proactive) mechanisms.

### 4.1 Security Evaluations using the HARMs with Security Metrics

Depending on the attack or defense methods, attack effort or defense efforts may vary. Therefore, quantifying the impact of attacks or defense can demonstrate the effort required for attacks or the strength of countermeasures. Security metrics are used with GSMs to present the security posture of networks in a quantitative manner based on a certain scale, which takes into account the impact of attacks or the effect of countermeasures. In our previous papers, we have adopted several security metrics to quantitatively assess the security of various types of networks with the HARMs. In addition, we have also developed several security metrics to represent the security posture of networks having different features (e.g., Enterprise, Cloud, IoT, SDN, *etc.*). Moreover, we have developed security metrics to evaluate the effectiveness of cyber-defenses including MTD and cyber-deceptions.

In the following section, we describe the major security metrics used to measure security and the effectiveness of defenses based on the different variants of HARMs. In Figure 3, we show a pictorial classification of the security metrics used with the HARMs.

#### 4.1.1 Metrics for measuring vulnerabilities and their relationships

Vulnerabilities from a networked system can be collected and analyzed using all the versions of HARMs based on the Common Vulnerability Scoring Systems (CVSS) metrics. Moreover, many attacks are performed based on multiple vulnerabilities or multi-step vulnerabilities. The metrics in this category measures the severity of vulnerabilities on a system based on the models in the lower layer of the HARM, such metrics include; (1) attack cost to measure the effort required to exploit a vulnerability, (2) attack success probability to measure the likelihood that a vulnerability will be exploited, including its dependency or relationship with other vulnerability, (3) vulnerability impact to measure the potential loss when an attacker exploit the vulnerability, *etc* [10, 36, 43]. In addition, we have used different path-based metrics (number of attack paths, shortest path, *etc*) in the HARMs to quantify attack scenarios based on vulnerabilities and their relationships with other vulnerabilities in the network.

#### 4.1.2 Metrics for measuring attacks scenarios

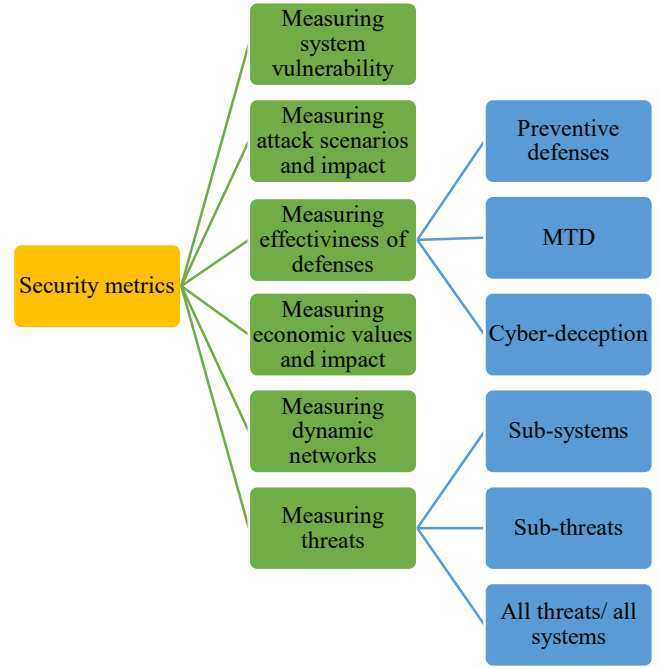

Figure 3: pictorial classification of security metrics used with the variants of HARMs.

This category of metrics measures the potential impact associated with the attack scenarios to achieved the attackers' goal. We group the metrics into two: probability-based and non-probability-based metrics. The probability-based metrics are computed based on the likelihood that an attack emerging, detected, prevented, *etc* taking into account node reachability information. We have used the probability-based metrics with HARM in several of our papers [10, 37, 38, 43, 44], and they include the probability of attack success on paths, the probability of an attacker interacting with a decoy [45], the probability that a node is connected with another node[46][52], etc. The non-probability-based metrics do not use probability values to measure impact or damages that an attacker may cause based on attack paths to achieving an attack goal. We have implemented many of these metrics and they include risk on attack paths, impact on attack paths, RoA on attack paths, *etc* [8][52].

#### 4.1.3 Metrics for measuring the effectiveness of defenses

Defenses deployed to networks need to be evaluated to understand their effectiveness. The defense metrics aim to measure the effectiveness of defense mechanisms placed in a system. We discuss the defense metrics we have developed or adopted with the HARMs in terms of preventive, reactive, and proactive defense mechanisms. The effectiveness of preventive defense such as security patches, firewalls, is evaluated based on the HARMs using the metrics for measuring vulnerabilities, metrics for measuring the impact of attacks, *etc*. Moreover, dynamic security metrics are developed to evaluate the effectiveness of MTD for the SDN and IoT networks.

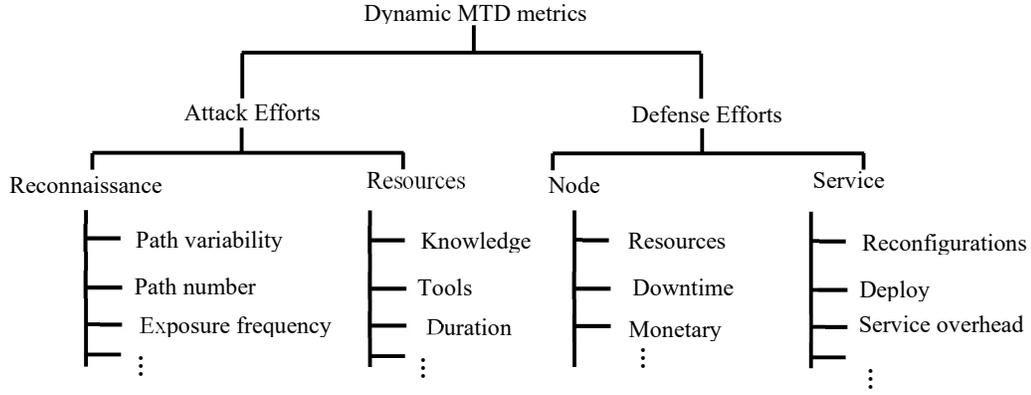

Figure 4: Dynamic metrics used with T-HARM to evaluate effectiveness of MTD techniques in SDN

To compare MTD techniques, relevant attack, and defense effort, we developed and classified the dynamic metrics to evaluate the effectiveness of MTD into two main categories including attack efforts and defense efforts. These metrics take into account the security changes introduced by MTD techniques, then measure their effectiveness based on the observed changes. In Figure 4, we show the classification of the dynamic MTD metrics used with T-HARM for the SDN. The attack effort metrics are categorized into two groups: reconnaissance and scanning. The reconnaissance metrics measure the effect of changes (MTD) with respect to the attacker gathering network information and the observed configuration changes. The resource metrics measure the properties of the attacker in terms of capabilities, tools, knowledge, and time taken to perform attacks when an MTD technique is deployed. On the other hand, the defense effort metrics measure the costs associated with deploying MTD techniques in the network, where the node metrics measure the cost/downtime with respect to changes (e.g., changes in terms of OS variant, edges, etc). Similarly, the service metrics measure the overhead incurring as a result of communication maintenance. The details of dynamic security metrics can be found in our previous work [47][53].

Furthermore, the HARM has been used to measure security, performance, and service availability in the IoT network as a result of deploying integrated proactive defense techniques consisting of cyber-deception (i.e., a decoy system) and moving target defense (i.e., network topology shuffling) [46]. In the [46], the following metrics are used with the HARM to evaluate the effectiveness of this integrated defense technique: the number of attack paths towards decoys, the meantime to a security failure, defense cost, and packet delivery ratio. Also, the HARM has been used with a Stochastic Reward Net to measure capacity-oriented availability and security (in terms of attack success probability) of enterprise networks under potential attacks before and after a security patch [48].

### 4.1.4 Metrics for economic values

Due to monetary constraints, not all defenses can be deployed. As a result, it is important to assess the cyber-defenses based on their economic profitability and deterrent effects on cyber-attacks in order to select and prioritize defenses. To calculate the economic benefit of defenses before deploying them, we have implemented several economic metrics in HARM based on cost models. In particular, in our paper [49], we have measured loss expectancy, the benefit of security, the cost of security, Return on Security Investment, and Return on Attack along with the HARM. In addition, in [47], we have used defense costs with respect to the costs associated with software-based diversity in MTD. Moreover, we have implemented the economic metrics, security metrics with a multi-objective algorithm to analyze defenses and select optimal ones to deploy based on their effectiveness and monetary costs [50].

### 4.1.5 Metrics for measuring the security of dynamic networks

Emerging networking technologies are flexible, elastic, and able to change their network configurations over time. This introduces an unknown security posture at different times. Hence making it difficult to understand the security posture or to perform accurate security analysis. To solve this problem, we utilize the T-HARM to develop new security metrics named stateless security risk [51]. The main idea of the stateless metrics is to combine the security posture of network states at different times to provide a security overview. This metric provides the network state-independent view of the security posture and assesses the security evolution of network states.

### 4.1.6 Metrics for measuring threats

To assess the risk associated with different types of threats for the cloud and SDN, this category measures the impact of threats based on the collected vulnerabilities. According to Microsoft's STRIDE threat modeling framework, the threat specific risk metric is developed to analyze the impact of specific threats for the cloud networks based on T-HARM [44]. This metric takes into account different categories of threats (e.g., spoofing, tampering, *etc.*). Moreover, the MV-HARM is used for the SDN to measure the impact of threats in terms of vulnerabilities, attack scenarios, and defenses [39][54].

4.2 *Practical Applications of the HARM in different domains*

In this section, we discuss the application of the HARMs in different domains, including IoT, SDNs, Cloud-based web services, and modeling MTD techniques and evaluation.

*Application in IoT networks*: IoT is characterized by a large number of heterogeneous and resource constraint devices, in which they continuously pose new security issues. Hence, modeling the security of IoT is a challenging task. HARM was employed to automate security assessments for the IoT [10]. Potential attack paths are captured to depict sequences of attack actions and security metrics are developed to evaluate the security level of the IoT. Moreover, due to constrained resources and limited computation capabilities, several proactive defense mechanisms have been proposed, including network topology shuffling-based MTD and cyber-deception which are evaluated based on the HARM. Furthermore, the evaluation results from the HARM have been taken as input into optimization algorithms to compute optimal defense deployment for the IoT.

*Application in Cloud networks*: Cloud computing offers highly scalable and dynamic features, as well as different privileged boundaries between stakeholders such that it is challenging to assess their security. A tool named *Cloudsafe* [9] was developed and deployed on the Amazon Elastic Compute Cloud platform to assess and evaluate their security. The CloudSafe tool gathers various tools including the HARM (as the evaluation engine) to automate the security assessment process. Using this tool, cloud service providers and individuals can generate a security report to understand the security posture of their cloud systems. In addition, the CloudSafe provides a way to pre-evaluate countermeasures before they are deployed. Moreover, another tool named '*ThreatRiskEvaluator*' was developed for threat-specific risk analysis based on vulnerability information, the probability of an attack, as well as client-specific security requirements [44]. This tool will allow cloud providers to make fine-grained decisions for selecting countermeasures to meet user requirements.

*Application in modeling and evaluating MTD techniques*: MTD is a defense strategy that continuously creates uncertainty for cyber attackers by dynamically changing the attack surface. Many MTD techniques have been proposed in the past to thwart cyberattacks (e.g., Shuffle, Diversity, and Redundancy). The first challenge to measuring the effectiveness of these defense techniques is "how to capture and model" the dynamic attributes of the networks as a result of the deployment of MTD techniques. Different variants of HARMs are employed to capture and model the changes introduced by MTD and measure its effectiveness.

## 5 FUTURE RESEARCH DIRECTION

A lot has been done in the area of graphical security modeling-based security analysis, there are still many problems that need to be addressed more effectively. We suggest future research directions in the following areas:

- **Adaptability**: Modern networks allow their components to change frequently, causing the frequent change of security posture and the effectiveness of security countermeasures. The temporal-GSMs are proposed to capture the security changes every time *t* and to evaluate the effectiveness of defenses. In practice, networks can be even more dynamic with thousands of state changes over time, where the temporal GSM is not able to tackle. To take into account the dynamic nature of modern networks, a more robust method is required for effective security evaluation.
- **Scalability**: Hierarchical modeling provides a way to address the scalability problem of GSMs. However, it is still challenging to generate and evaluate the security of large-scale networks (e.g., IoT) with possibly every node as an entry point and target. The efficient design and implementation of the models can be further explored.
- **Lack of empirical data**: GSMs require connectivity and vulnerability information of the nodes as input. Due to the inclusion of cyber-physical systems and emerging IoT, the lack of empirical data becomes a big limitation in security evaluation via GSMs. Approaches to tackle the absence of empirical data in GSMs can be further explored.

## 6 CONCLUSION

In this paper, we have surveyed the development and application of graphical security models. In particular, we have discussed the evolution of security models from AGs, ATs, DTs, and to ACTs. We have categorized the security models into graph-based, tree-based, and hybrid and provided definitions. Furthermore, we have provided a comprehensive survey of the hierarchical security model and its applications in various domains to enable users to comprehend how each HARM variant can be applied. By doing so, users can have a better understanding of which version of the HARM can effectively address their security needs and concerns. Besides, we have summarized the pros and cons of each variant of the hierarchical security model. In addition, we have classified the various security metrics used with the HARMs based on what they measure and their applicability. Finally, we pointed out future research challenges and directions based on graphical security models and their applications.

## BIOGRAPHIES

Simon Yusuf Enoch, PhD
School of Information Technology and Electrical Engineering,
The University of Queensland,
Brisbane,  QLD 4072, Australia

e-mail: sey19@uclive.ac.nz

Simon Yusuf Enoch received a Ph.D. degree in Computer Science from the University of Canterbury (UC), New Zealand in 2018. He is a Postdoctoral Research Fellow with the School of Information Technology and Electrical Engineering, The University of Queensland (UQ), Australia, where he is being mentored by Assoc. Prof. Dong Seong Kim. Prior to UQ, Dr. Enochson was a Research Assistant with the Cybersecurity Research Lab. at UC, New Zealand from 2017 to 2019. He is also working as a Lecturer in Computer Science with the Federal University Kashere, Gombe, Nigeria. Dr. Enochson has published papers in reputable Conferences and top-tier Journals. His research interests include cyber-attacks & defense automation, security modeling, and analysis of computers and networks including moving target defense.

Mengmeng Ge, PhD
School of Information Technology,
Deakin University, Geelong,
Australia

e-mail: mengmeng.ge@deakin.edu.au

Mengmeng Ge is a Lecturer in Cyber Security at Deakin University. She received her Ph.D. degree in Computer Science from the University of Canterbury, New Zealand, under the topic of graphical security modeling and assessment for the Internet of Things. Her research interests are graphical security modeling, the Internet of Things, software-defined networking, and data science in cybersecurity.

Jin B. Hong, PhD
Department of Computer Science and Software Engineering,
University of Western Australia,
 Australia

e-mail: jin.hong@uwa.edu.au

Jin B. Hong received a Ph.D. degree in computer science from the University of Canterbury, New Zealand. He is currently a Lecturer with the Department of Computer Science and Software Engineering, The University of Western Australia, Australia. His research interests include security modeling and analysis of computers and networks including cloud computing, SDN, and the IoT, and moving target defense.

Dong Seong Kim, PhD
School of Information Technology and Electrical Engineering,
The University of Queensland,
Brisbane,  QLD 4072, Australia

e-mail: dan.kim@uq.edu.au

Dong Seong Kim is an Associate Professor at the University of Queensland (UQ), Brisbane, Australia. Prior to UQ, he led the Cybersecurity Lab. at the University of Canterbury (UC), Christchurch, New Zealand from August 2011 to Dec 2018. He was a Senior Lecturer in Cyber Security in the Department of Computer Science and Software Engineering at the UC.  He was a visiting scholar at the University of Maryland, College Park, Maryland in the US during the year 2007. He received a Ph.D. degree in Computer Engineering from the Korea Aerospace University in February 2008. From 2008 to 2011, he was a postdoc at Duke University, Durham, North Carolina in the US. His research interests are in cybersecurity modeling and analysis for various systems and networks including IoT/CPS, Cloud, SDN.